\documentclass{article}

\usepackage{arxiv}

\usepackage[utf8]{inputenc} % allow utf-8 input
\usepackage{float}
\usepackage[T1]{fontenc}    % use 8-bit T1 fonts
\usepackage{hyperref}       % hyperlinks
\usepackage{url}            % simple URL typesetting
\usepackage{booktabs}       % professional-quality tables
\usepackage{amsfonts}       % blackboard math symbols
\usepackage{nicefrac}       % compact symbols for 1/2, etc.
\usepackage{microtype}      % microtypography
\usepackage{graphicx}
\usepackage{doi}
\usepackage{subcaption}
\usepackage[sorting=none]{biblatex} 
\usepackage{framed}
\usepackage{caption}

\title{EEG-GPT: Exploring Capabilities of Large Language Models for EEG Classification and Interpretation}

\author{ \href{https://orcid.org/0000-0000-0000-0000}{\includegraphics[scale=0.06]{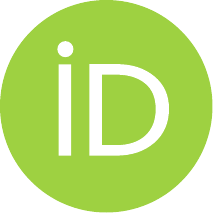}\hspace{1mm}Jonathan W. Kim, BS} \\
	Carle Illinois College of Medicine\\
	  Urbana-Champaign, IL  \\
	\texttt{jwk7@illinois.edu} \\
	\And
	\href{https://orcid.org/0000-0000-0000-0000}{\includegraphics[scale=0.06]{orcid.pdf}\hspace{1mm}Ahmed Alaa, PhD} \\
	Department of EECS\\
	UC Berkeley\\
	  Berkeley, CA  \\
        \And
	\href{https://orcid.org/0000-0000-0000-0000}{\includegraphics[scale=0.06]{orcid.pdf}\hspace{1mm}Danilo Bernardo, MD} \\
	Department of Neurology\\
	UC San Fracisco\\
	  San Francisco, CA  \\
	\texttt{dbernardoj@gmail.com} \\
}

\renewcommand{\shorttitle}{\textit{arXiv} A Preprint}

\hypersetup{
pdftitle={EEG-GPT: Exploring Capabilities of Large Language Models for EEG Classification and Interpretation},
pdfsubject={q-bio.NC, q-bio.QM},
pdfauthor={Jonathan W. ~Kim, BS, Ahmed ~Alaa, Danilo ~Bernardo},
pdfkeywords={Large language models, EEG},
}

\begin{document}
\maketitle

% TODO trim abstract down!!
\begin{abstract} 
\textbf{Rationale}
In conventional machine learning (ML) approaches applied to electroencephalography (EEG), this is often a limited focus, isolating specific brain activities occurring across disparate temporal scales (from transient spikes in milliseconds to seizures lasting minutes) and spatial scales (from localized high-frequency oscillations to global sleep activity). This siloed approach limits the development EEG ML models that exhibit multi-scale electrophysiological understanding and classification capabilities. Moreover, typical ML EEG approaches utilize black-box approaches, limiting their interpretability and trustworthiness in clinical contexts. Thus, we propose EEG-GPT, a unifying approach to EEG classification that leverages advances in large language models (LLM). EEG-GPT achieves excellent performance comparable to current state-of-the-art deep learning methods in classifying normal from abnormal EEG in a few-shot learning paradigm utilizing only 2\% of training data. Furthermore, it offers the distinct advantages of providing intermediate reasoning steps and coordinating specialist EEG tools across multiple scales in its operation, offering transparent and interpretable step-by-step verification, thereby promoting trustworthiness in clinical contexts.

\end{abstract}

% keywords can be removed

\section{Introduction}
Large language models (LLMs) such as ChatGPT have garnered substantial attention in the media and among the machine learning (ML) community. LLMs represent a pivotal paradigm shift in artificial intelligence (AI), consisting of transformer architectures substantially larger in scale compared to their predecessors, such as Recurrent Neural Networks (RNNs) and Long Short-Term Memory (LSTM) networks \cite{vaswani2017attention}, and leverage internet-scale text corpora, thus excelling not only on text completion tasks but demonstrating emergent capabilities in rudimentary language reasoning \cite{lightman2023lets, nye2021work}.

LLMs display several features conducive to the small data regime present in most EEG datasets, where the largest datasets typically have on the order of only thousands of EEGs. Primarily, LLMs have the capability to perform few- and even zero-shot learning \cite{wang2020fewshot}. Recent research has investigated how LLMs can perform few-shot learning in domains ranging from cancer drug synergy prediction to cardiac signal analysis \cite{liu2023large, li2023cancergpt}. Other work has demonstrated the ability of LLMs to outperform experts in annotating political Twitter messages with zero-shot learning \cite{törnberg2023chatgpt4}. Additionally, previous work has shown that transformer architectures are capable of utilizing in-context learning for zero-shot tasks -- in other words, utilizing information provided in the prompt in order to yield better performance on various tasks \cite{NEURIPS2022_c529dba0}.

\begin{figure}[H]
\centering
\includegraphics[width=0.85\textwidth]{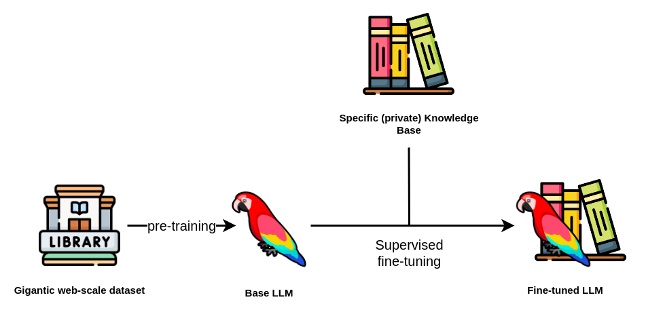}
\caption{Diagram depicting the process of fine-tuning large language models \cite{bratanic2023diagram}}\label{intro_finetuning}
\end{figure}

LLMs also appear to have the ability to plan and carry out intermediate reasoning steps when asked to solve complex problems. Lightman et al. investigate this property by training an LLM to solve complex competition-level math problems in a stepwise manner that is transparent and human-interpretable \cite{lightman2023lets}. Previous work has investigated using prompting strategies such as Chain of Thought \cite{nye2021work}, which have been shown to improve the ability of LLMs to perform multi-step computations. More recently, Yao et al. developed a framework called Tree of Thoughts, an extension of Chain of Thought approaches which constructs a decision tree to be explored by an LLM in order to facilitate solving complex problems requiring strategic lookahead and backtracking \cite{yang2023mmreact}. 

Finally, LLMs appear to be capable of synergizing inputs from other computational “experts” as a human might. In other words, when asked to solve a complex problem, if given access to other specialized computational tools, LLMs exhibit the capability to coordinate passing appropriate inputs to those specialist tools then synergizing their outputs to come to a result. Yang et al. developed a system paradigm known as MM-REACT which integrated ChatGPT with a pool of computational "experts" in order to perform vision-based tasks in a zero-shot regime \cite{yang2023mmreact}. 

Previous work has investigated the application of LLMs to EEG-related tasks. Cui et al. report NeuroGPT, an LLM-based framework used to generate and predict embeddings of EEG signal \cite{cui2023neurogpt}. They demonstrate that this framework shows improved performance on the task of classifying motor imagery from EEG signal relative to other machine learning modalities. Other work includes the BENDR framework by Kostas et al., which applied a transformer architecture towards the task of EEG-based sleep stage classification \cite{kostas2021bendr}.

Human EEG classification and interpretation is highly subjective, with prior studies in interrater reliability demonstrating widely variable Cohen’s kappa ($\kappa_c$) ranging from 0.3 to 0.7 amongst trained epileptologists\cite{grant2014eeg}. Moreover, recent evaluations of current state-of-the-art (SOTA) EEG interpretation systems such as Persyst which utilizes deep neural networks have demonstrated subpar performance, with significant rates of false negatives and false positives raising questions about the current utility of DL EEG systems in clinical practice\cite{ganguly2022seizure}. Here, we aim to evaluate the potential of LLMs to aid clinicians in the tasks of EEG classification and interpretation. We investigate whether an LLM-based approach may offer advantages over current DL-based EEG interpretation and classification methods in performance as well as in transparency.  

\section{Methods}
We utilized two approaches to explore the capabilities of LLMs as they apply to clinical EEG tasks, specifically the task of classifying EEG as normal or abnormal. In both, we utilize the Temple University Hospital Abnormal Corpus, which consists of a total of 1140 hours of EEG data collected from 2993 subjects, is roughly evenly balanced between normal and abnormal recordings, and is pre-split into train and evaluation sets for consistency of evaluation across different experiments \cite{lopez2017tuab}.

\subsection{Few- and zero-shot Learning}
\begin{figure}[H]
\centering
\includegraphics[width=0.85\textwidth]{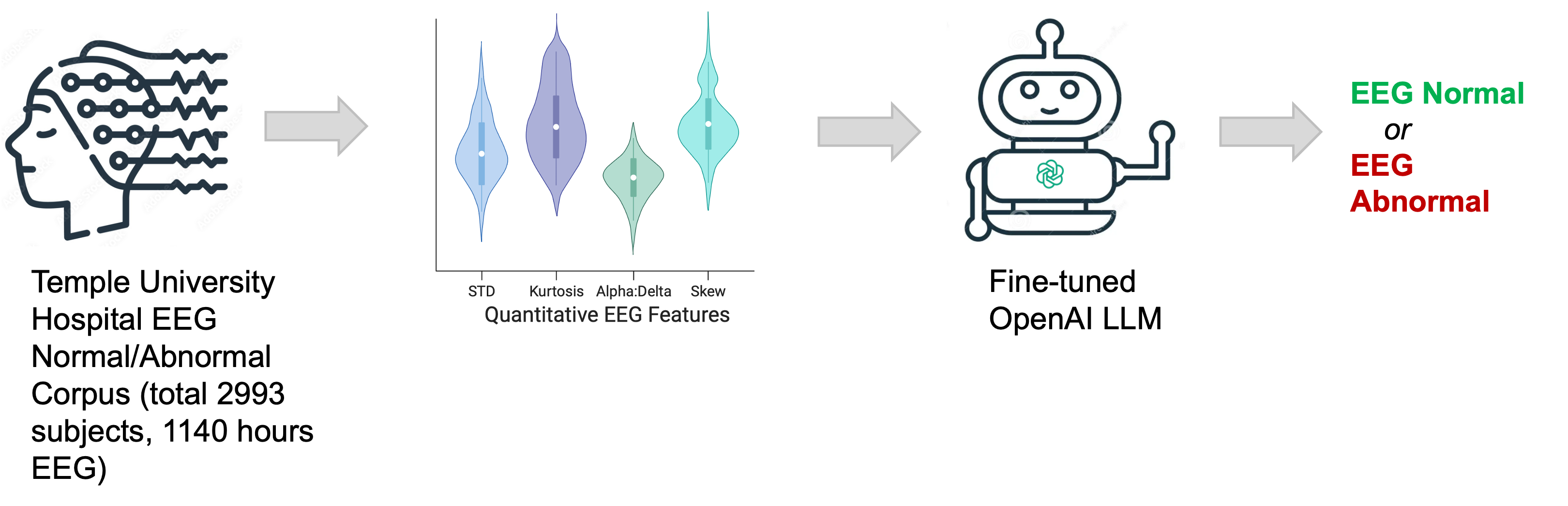}
\caption{Pipeline for few-shot experiment}\label{exp1_diagram}
\end{figure}

We hypothesize that, given a relatively small amount of training data, a fine-tuned LLM (from here on referred to as EEG-GPT) will be able to classify EEG as normal or abnormal at a high level of performance relative to other machine learning modalities. We also hypothesize that with zero-shot learning, the base LLM will be able to perform significantly better than chance on the normal / abnormal classification task. Our pipeline for this experiment is shown in Figure \ref{exp1_diagram}.

\paragraph{Feature selection and fine-tuning} 
For a given EEG file, we subdivide the file into non-overlapping 20-second epochs. For each epoch, we calculate per-channel features (the subset of channels used is shown in Table \ref{table:channels} and the set of features used is shown in Table \ref{table:feats}). The same set of features are calculated for each channel in order to yield a 30-feature (6 features times 5 channels) sample for each epoch. Each feature sample is labeled as normal or abnormal according to whether its parent file was labeled as normal or abnormal.

We then convert these features to a verbal representation and then use these verbal representations both to fine-tune and evaluate the da Vinci GPT-3 base LLM from OpenAI on the resulting feature set. More specifically, we use OpenAI's Completions API in order to fine-tune and evaluate EEG-GPT; to fine-tune an LLM using this API, we provide prompt and completion pair examples, with the prompt being a verbal representation of features, and the completion being the normal / abnormal label (an example prompt-completion pair is shown in Figure \ref{fig:examplepromptcompletion}). OpenAI's API then fine-tunes the LLM to learn to "complete" a prompt with the corresponding completion, in this case the prediction of whether a given file is normal or abnormal. 

\begin{minipage}{.4\textwidth}
    \centering
        \begin{tabular}{ |c| }
            \hline
            Features calculated (over 20-second epoch) \\
            \hline
            90th percentile of voltage amplitudes \\
            standard deviation \\ 
            kurtosis \\ 
            alpha:delta power ratio \\
            theta:alpha power ratio \\ 
            delta:theta power ratio \\
            \hline
        \end{tabular}
    \captionof{table}{Features calculated per channel}
    \label{table:feats}
\end{minipage}%
\begin{minipage}{.5\textwidth}
    \centering
        \begin{tabular}{ |c| }
            \hline
            Channels used for feature calculation \\
            \hline
            Cz \\
            T5 \\
            T6 \\
            O1 \\ 
            O2 \\
            \hline
        \end{tabular}
    \captionof{table}{Channels used for feature calculation}
    \label{table:channels}
\end{minipage}

\begingroup
    \begin{framed}
            \begin{verbatim}
    {"prompt":
        " Quantitative EEG: In a 20 second period, 
            at channel Cz:[
                the 90th percentile of voltage amplitudes = 0.35 microvolts, 
                standard deviation = 0.27, 
                kurtosis = 0.33, 
                alpha:delta power ratio = 0.22, 
                theta:alpha power ratio = 9.11, 
                delta:theta power ratio = 0.49]; 
            at channel T5:[
                the 90th percentile of voltage amplitudes = 0.10 microvolts,
                standard deviation = 0.09,
                kurtosis = 1.53,
                alpha:delta power ratio = 0.13,
                theta:alpha power ratio = 2.43,
                delta:theta power ratio = 3.13]; 
            at channel T6:[
                the 90th percentile of voltage amplitudes = 0.18 microvolts,
                standard deviation = 0.20, 
                kurtosis = 9.47, 
                alpha:delta power ratio = 0.09, 
                theta:alpha power ratio = 3.80, 
                delta:theta power ratio = 2.95]; 
            at channel O1:[
                the 90th percentile of voltage amplitudes = 0.34 microvolts, 
                standard deviation = 0.28, 
                kurtosis = 0.11, 
                alpha:delta power ratio = 0.82, 
                theta:alpha power ratio = 1.33, 
                delta:theta power ratio = 0.92]; 
            at channel O2:[
                the 90th percentile of voltage amplitudes = 0.30 microvolts, 
                standard deviation = 0.26, 
                kurtosis = 2.47, 
                alpha:delta power ratio = 0.39, 
                theta:alpha power ratio = 1.35, 
                delta:theta power ratio = 1.89];.
    
            Cumulative Effect Category:",
    
    "completion":" normal"}
            \end{verbatim}
            \label{fig:examplepromptcompletion}
    \end{framed}
    \captionof{figure}{Example prompt-completion pair (formatted for readability)}
\endgroup

\paragraph{Few- and zero-shot learning} Using our derived feature set, we explore EEG-GPT's performance on the normal/abnormal classification task in a few-shot context, and compare its performance against traditional machine learning approaches as well as against more recent deep-learning based methods. We also explore EEG-GPT's performance in a zero-shot context -- in other words, we evaluate how EEG-GPT performs on the normal/abnormal task with no fine-tuning whatsoever, both with and without in-context learning. 
%TODO put example of in-context learning in here?

\subsection{Evaluation of reasoning capability in EEG tool usage}
\begin{figure}[H]
\centering
\includegraphics[width=0.5\textwidth]{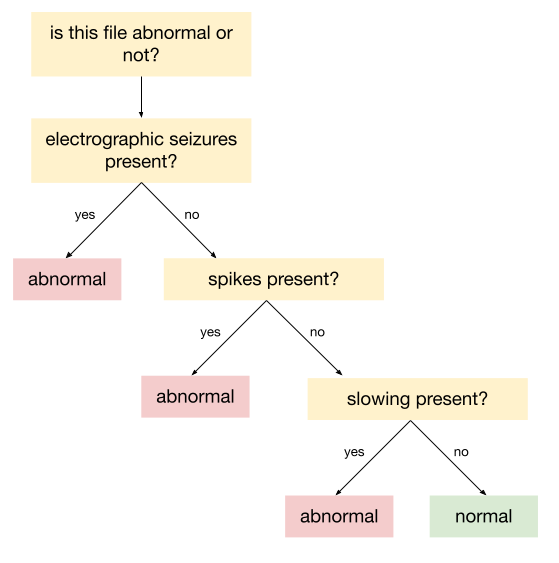}
\caption{Simplified diagram of clinical epileptologist workflow}\label{exp2_diagram}
\end{figure}

We also hypothesize that an LLM-based framework will be able to effectively utilize specialized software tools to classify EEG as abnormal or normal, emulating how EEG readers systematically evaluate for signs of seizure, spikes, and/or slowing when making a classification. To develop this framework, we consider how clinical epileptologists analyze a given EEG file for abnormalities.We conceptualize this process as a decision tree, which is depicted in Figure \ref{exp2_diagram}. In this conceptualization, the epileptologist may first evaluate the file for evidence of seizures, then check for spikes, then check for signs of slowing -- at any point, if there is strong evidence that any of these phenomena are occurring, the epileptologist can confidently claim that the file is abnormal. This is an overly simplified view of how clinical epileptologists analyze EEG, but is appropriate for initial investigations into how LLMs might utilize a similar step-wise, decision-tree-based process when coming to an analytical decision of a file.

\paragraph{Tree of Thought} In order to adapt LLMs to solve complex problems requiring exploration and strategic lookahead, Yao et al. developed the Tree-of-Thought framework, which allows LLMs to explore and backtrack along a decision-tree style exploration space when solving a complex problem \cite{yao2023tree}. This approach maps nicely onto our simplified decision tree depicted in Figure \ref{exp2_diagram}, in which a clinical epileptologist or LLM will explore the decision tree space in order to classify a file as normal or abnormal.
\paragraph{Specialized software tools} In order to enable exploration of the decision tree we provide an agentic LLM with access to three software tools: an automated seizure detection model, an automated spike detection model, and an automated qEEG feature comparison tool. When given a particular EEG file to evaluate, the framework will explore the decision tree space, using the automated software tools available to it at each branching point, deciding whether or not to proceed based on the input from each software tool. These software tools are described in greater detail below.
\paragraph{Automated seizure detection model} Using the Temple University Hospital seizure corpus \cite{obeid2016tuhsz}, we trained an automated EEG-based seizure detection model using a convolutional neural network architecture. We ensure that there is no overlap between the subjects used to train the seizure detection model and the subjects used to evaluate the overall framework. This model takes in 10-second epochs and returns a yes/no decision as to whether a given epoch contains seizure activity or not. The model's architecture and performance on a held-out set (pulled from a set of subjects that does not overlap with the subjects used to evaluate the overall framework) are depicted below. 
\paragraph{Automated spike detection model}We use an automated spike detection algorithm based on Esteller's work investigating line length as a computational biomarker for EEG spikes \cite{esteller2001line}.
\paragraph{qEEG feature comparison tool}As a final check for abnormality, we provide the framework with a qEEG feature comparison tool. For this tool, we calculate a reference of age-based normative ranges for a subset of qEEG features (statistical moments and power ratios). Then, when given a particular EEG file, this tool calculates the set of qEEG features and calculates a cosine similarity score to the age-matched normative reference. If the cosine similarity score is higher than a particular threshold, it will return a result of "similar to a reference of normal files," and vice versa. This tool is also capable of providing a confidence score associated with its prediction based on calculating positive and negative predictive values derived from a held-out set.
%TODO diagram for qEEG comparison tool?
%TODO do i need to mention pinecone specifically?

\section{Results}
\subsection{EEG-GPT demonstrates few- and zero-shot learning proficiency}
We evaluate EEG-GPT in both few- and zero-shot contexts against both traditional machine learning approaches and more recent deep-learning based approaches. For the comparison against traditional machine learning approaches, we utilize the same set of training features in order to facilitate appropriate comparisons between the performance profiles of each. Comparisons against deep-learning based approaches utilize the performance metrics reported in their respective papers given that their deep learning architectures utilize the raw EEG inputs instead of the limited feature set used by EEG-GPT, and therefore their models would not be able to utilize the same feature set. 

\begin{figure}[H]
\centering
\includegraphics[width=0.65\textwidth]{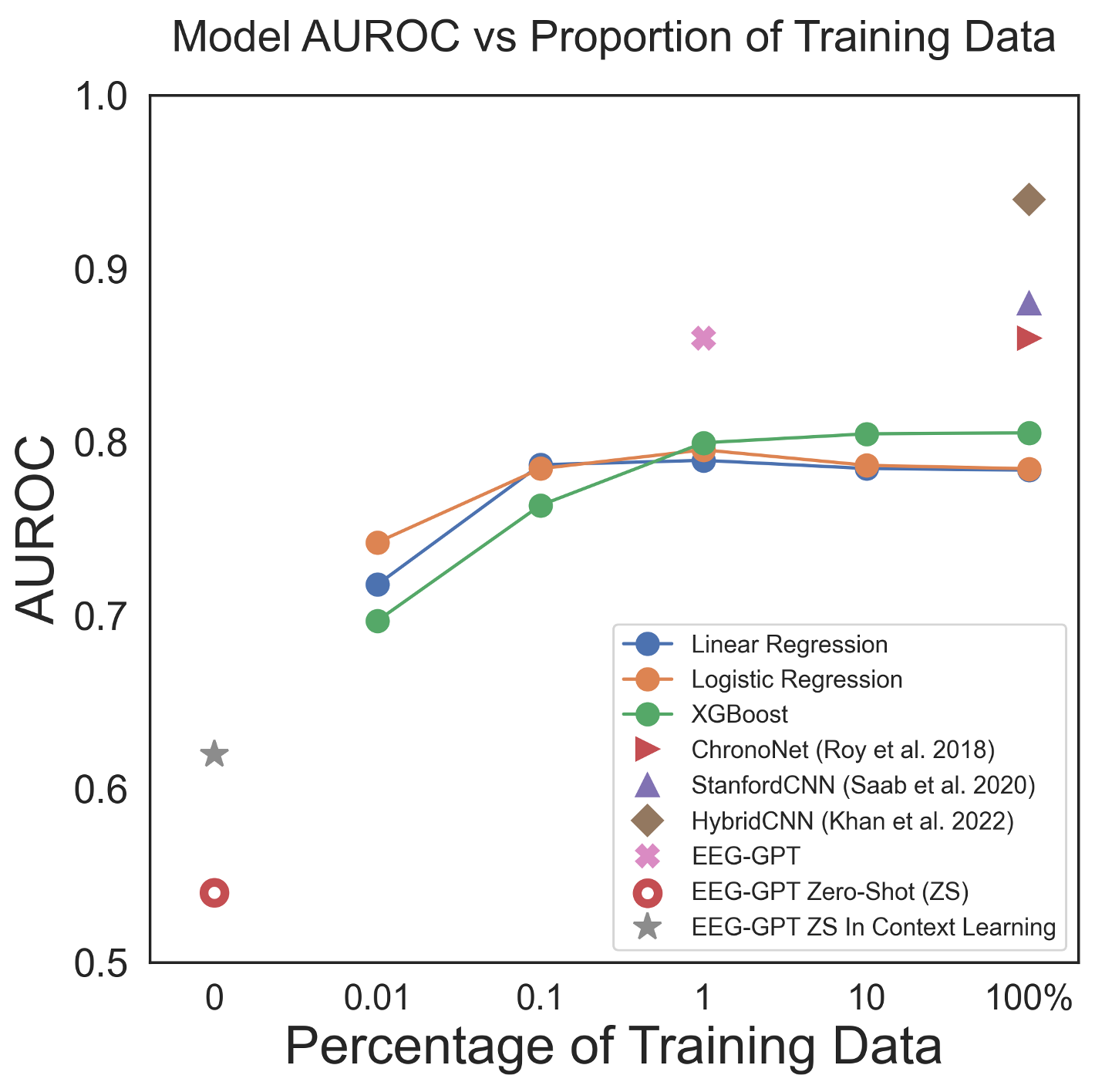}
\caption{Resulting AUROCs of normal/abnormal classification task, plotted against proportion of training data used to fit model}\label{exp1_results}
\end{figure}

When trained on 2\% of available data, EEG-GPT achieves an AUROC of 0.86, as shown in Figure \ref{exp1_results} (EEG-GPT's performance depicted with the pink "X"). This is better than traditional machine learning approaches (linear regression, logistic regression, and XGBoost), which improve performance when given increasing amounts of training data, but plateau at an AUROC of roughly 0.8. In the zero-shot context, EEG-GPT's performance remains better than chance, and improves to an AUROC of 0.63 when provided with in-context learning. When trained on 2\% of available train data, EEG-GPT even matches the performance of a deep-learning based approach trained on all available data \cite{roy2019chrononet}, although it is still outperformed by more recent approaches such as those by Saab et al. and Khan et al. \cite{saab2020weak, khan2022human}.

\subsection{EEG-GPT navigates EEG tool usage with tree-of-thought reasoning }
\begin{figure}[H]
\centering
\includegraphics[width=0.65\textwidth]{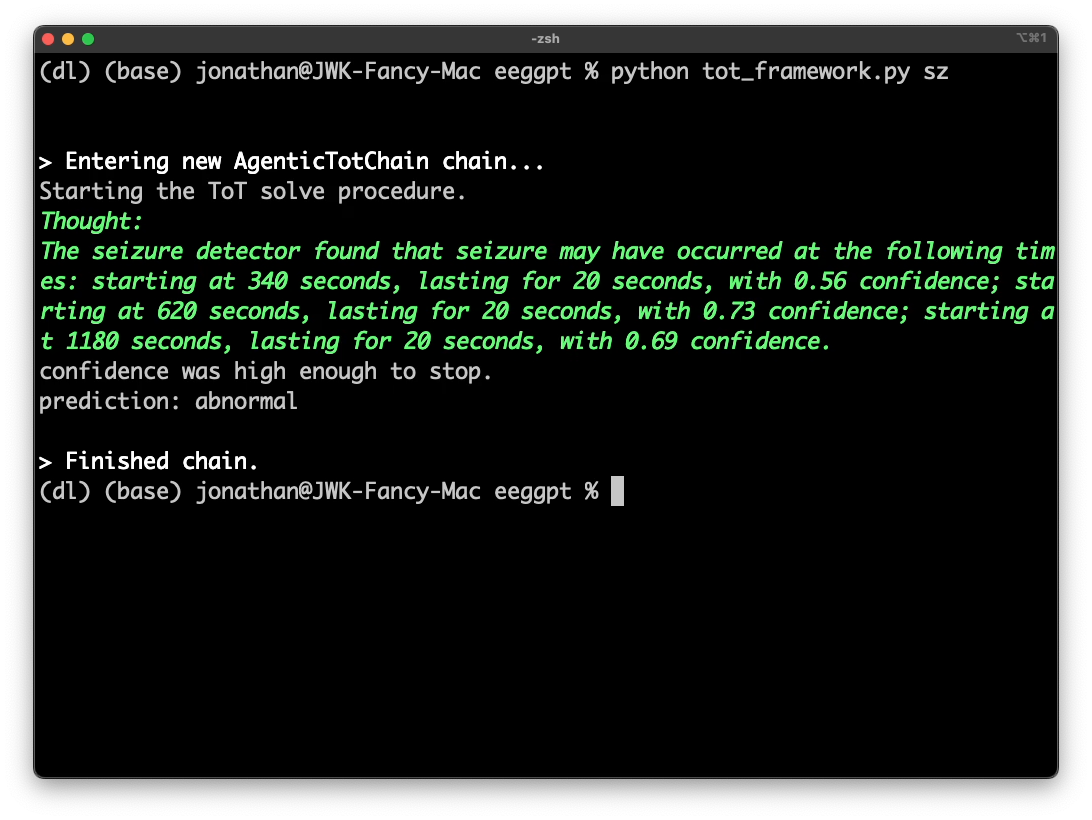}
\caption{Framework's analysis of EEG file known to contain seizure}\label{exp2_demosz}
\end{figure}

We present two exemplars of the tree-of-thought approach here. In the first demo, shown in Figure \ref{exp2_demosz}, we pass an EEG file which is known to contain seizures to our framework. Stepping through the solving procedure we see that the automated seizure detector immediately detects a seizure, indicating that the file is abnormal and allowing for early stoppage of the solving procedure.

\begin{figure}[H]
\centering
\includegraphics[width=0.65\textwidth]{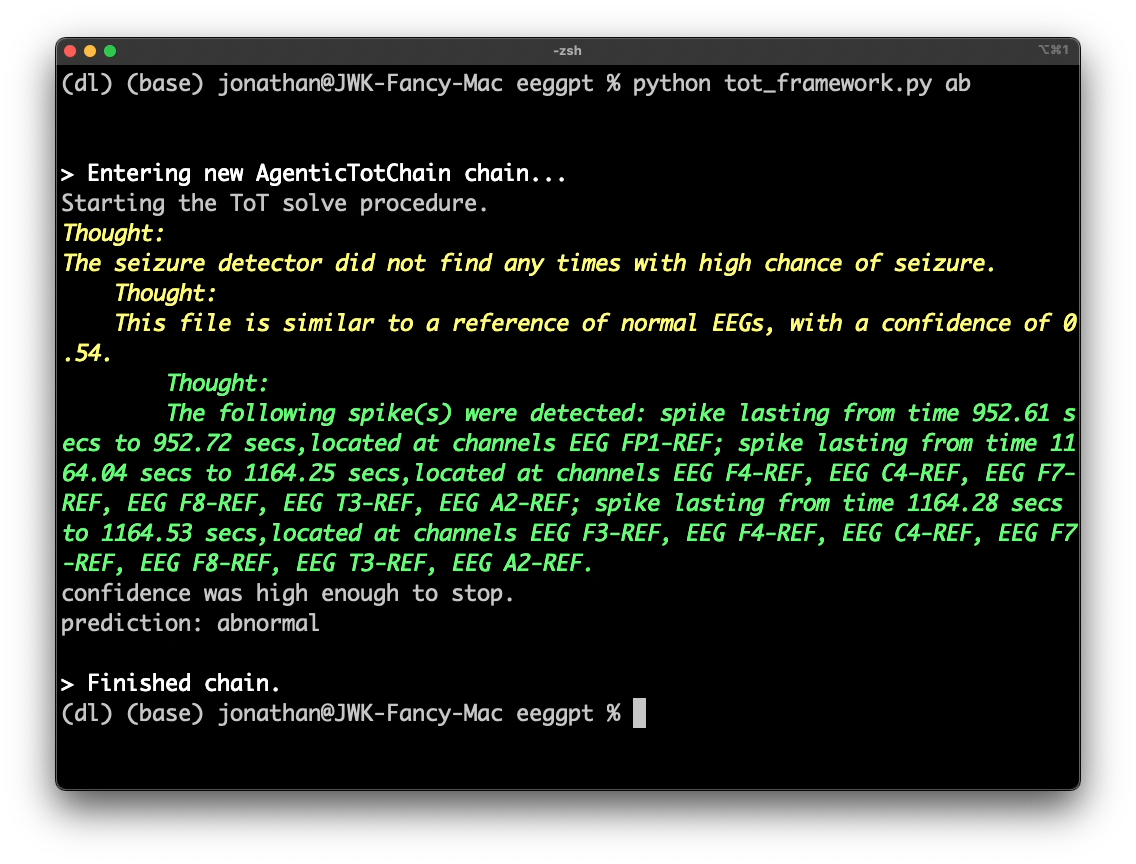}
\caption{Framework's analysis of abnormal EEG file known to be seizure-free}\label{exp2_demoab}
\end{figure}

In the second demo, shown in Figure \ref{exp2_demoab}, we pass an EEG file which is known to be abnormal yet not contain seizures to the framework. Stepping through the solve procedure, we see that the seizure detector does not find any seizures -- the framework then decides to evaluate the file with another tool, the qEEG feature comparison tool. The framework then finds that the file is similar to a reference of normal EEGs, with a confidence measure of 0.54 provided by the qEEG tool itself. Presumably given the low confidence, the framework decides it needs still more information before classifying this file, and proceeds to use the automated spike detection tool -- which does indeed detect spikes, indicating that this file is abnormal. At this point the framework decides it has enough information and confidence to make a final classification, and halts execution.

\section{Discussion \& Future Work}
In this work we demonstrate the proficiency of EEG-GPT within a few-shot learning paradigm for EEG interpretation and classification tasks. In addition, we demonstrate a pilot exploration of EEG-GPT's reasoning ability in its capacity to navigate usage of specialist EEG tools across multiple temporal scales in a step-wise, transparent fashion. For the normal/abnormal classification task, EEG-GPT demonstrates performance that surpasses traditional machine learning methods at all amounts of training data. EEG-GPT also matches at least one previous deep-learning based approach, even when trained on 50 times less data. It is also worth noting that while all the deep learning approaches evaluated (ChronoNet, StanfordCNN, and HybridCNN) outperform or match EEG-GPT, these approaches are trained on all available training data, and utilize the raw EEG signal, which is a much richer feature space than the relatively limited set of features used by EEG-GPT. EEG-GPT's few- and zero-shot learning abilities may be partially explained by the likely presence of normative EEG data in the large data corpus used to train the base LLM. 

Clinical medicine has traditionally been relatively slow to adopt machine learning techniques for clinical practice due to various ethico-legal reasons, as well as the fact that many state-of-the-art machine learning models function as "black boxes," with very little insight into how models come to a particular prediction or decision. By utilizing structured reasoning strategies such as tree-of-thought, LLMs are capable of demonstrating logical flows in their problem-solving processes. However, a significant limitation of current LLMs is their propensity to generate non-factual or  "hallucinated" information\cite{huang2023survey}, which limits their reliability and trustworthiness in high-stakes, clinical settings. To address this limitation, step-wise verifiability of clinical LLM systems coupled with a human in the loop or human oversight emerges as a strategy to identify LLM errors or hallucinations. This explainable AI (XAI) approach allows for monitoring and rectification of LLM system outputs, thereby mitigating the risks associated with hallucinations and easing the path to clinical adoption of AI. 

Future investigation is warranted to develop and validate the potential of LLM systems for EEG interpretation and classification. Future work would involve further exploration of the clinical EEG feature space in order to further optimize performance. Given our results which seem to indicate LLMs' usefulness in performing tasks with relatively small amounts of training data, there is reason to hypothesize that LLMs may be particularly useful for rare disease applications, where training data is relatively expensive or difficult to obtain. Further study is needed in evaluating the reasoning capabilities, such as in tree-of-thought, to identify potential failure modes such as "hallucinations". In addition, as current LLMs may have poorer performance with more complex reasoning tasks\cite{wang2023hypothesis}, more work is necessary to appraise the validity of LLM logical reasoning in EEG interpretation and classification problems. Further investigation would also involve continuing to add specialized software tools to the framework and further building out the system's ability to appropriately recognize edge cases that may merit specialized software tools. 

LLMs represent a promising approach for EEG interpretation and classification, and their potential to demonstrate emergent AI-reasoning abilities warrants further development and investigation. However, it is crucial that future work recognize limitations of LLMs' nascent "reasoning" capabilities especially their tendency to hallucinate. Consequently, a step-by-step verifiability with human oversight is highly recommended, as it enhances the reliability and interpretability of LLMs in clinical applications, aligning with the crucial need for trust and accountability in medical decision-making.

\printbibliography

@ARTICLE{kostas2021bendr,
    AUTHOR={Kostas, Demetres and Aroca-Ouellette, Stéphane and Rudzicz, Frank},   
    TITLE={BENDR: Using Transformers and a Contrastive Self-Supervised Learning Task to Learn From Massive Amounts of EEG Data},      
    JOURNAL={Frontiers in Human Neuroscience},      
    VOLUME={15},           
    YEAR={2021},      
    URL={https://www.frontiersin.org/articles/10.3389/fnhum.2021.653659},       
    DOI={10.3389/fnhum.2021.653659},      
    ISSN={1662-5161}
}

@misc{cui2023neurogpt,
      title={Neuro-GPT: Developing A Foundation Model for EEG}, 
      author={Wenhui Cui and Woojae Jeong and Philipp Thölke and Takfarinas Medani and Karim Jerbi and Anand A. Joshi and Richard M. Leahy},
      year={2023},
      eprint={2311.03764},
      archivePrefix={arXiv},
      primaryClass={cs.LG}
}

@misc{törnberg2023chatgpt4,
      title={ChatGPT-4 Outperforms Experts and Crowd Workers in Annotating Political Twitter Messages with Zero-Shot Learning}, 
      author={Petter Törnberg},
      year={2023},
      eprint={2304.06588},
      archivePrefix={arXiv},
      primaryClass={cs.CL}
}

@inproceedings{NEURIPS2022_c529dba0,
 author = {Garg, Shivam and Tsipras, Dimitris and Liang, Percy S and Valiant, Gregory},
 booktitle = {Advances in Neural Information Processing Systems},
 editor = {S. Koyejo and S. Mohamed and A. Agarwal and D. Belgrave and K. Cho and A. Oh},
 pages = {30583--30598},
 publisher = {Curran Associates, Inc.},
 title = {What Can Transformers Learn In-Context? A Case Study of Simple Function Classes},
 url = {https://proceedings.neurips.cc/paper_files/paper/2022/file/c529dba08a146ea8d6cf715ae8930cbe-Paper-Conference.pdf},
 volume = {35},
 year = {2022}
}

@inproceedings{roy2019chrononet,
  title={ChronoNet: A deep recurrent neural network for abnormal EEG identification},
  author={Roy, Subhrajit and Kiral-Kornek, Isabell and Harrer, Stefan},
  booktitle={Artificial Intelligence in Medicine: 17th Conference on Artificial Intelligence in Medicine, AIME 2019, Poznan, Poland, June 26--29, 2019, Proceedings 17},
  pages={47--56},
  year={2019},
  organization={Springer}
}

@article{saab2020weak,
  title={Weak supervision as an efficient approach for automated seizure detection in electroencephalography},
  author={Saab, Khaled and Dunnmon, Jared and R{\'e}, Christopher and Rubin, Daniel and Lee-Messer, Christopher},
  journal={NPJ digital medicine},
  volume={3},
  number={1},
  pages={59},
  year={2020},
  publisher={Nature Publishing Group UK London}
}

@article{khan2022human,
  title={Human activity recognition via hybrid deep learning based model},
  author={Khan, Imran Ullah and Afzal, Sitara and Lee, Jong Weon},
  journal={Sensors},
  volume={22},
  number={1},
  pages={323},
  year={2022},
  publisher={MDPI}
}

@ARTICLE{obeid2016tuhsz,
    AUTHOR={Obeid, Iyad and Picone, Joseph},   
    TITLE={The Temple University Hospital EEG Data Corpus},      
    JOURNAL={Frontiers in Neuroscience},      
    VOLUME={10},           
    YEAR={2016},      
    URL={https://www.frontiersin.org/articles/10.3389/fnins.2016.00196},       
    DOI={10.3389/fnins.2016.00196},      
    ISSN={1662-453X}   
}

@online{bratanic2023diagram,
    title = {Knowledge Graphs \& LLMs: Fine-Tuning vs. Retrieval-Augmented Generation},
    author = {Tomaz Bratanic},
    year = {2023},
    urldate = {1/17/2024},
    url={https://neo4j.com/developer-blog/fine-tuning-retrieval-augmented-generation/}
}

@INPROCEEDINGS{esteller2001line,
  author={Esteller, R. and Echauz, J. and Tcheng, T. and Litt, B. and Pless, B.},
  booktitle={2001 Conference Proceedings of the 23rd Annual International Conference of the IEEE Engineering in Medicine and Biology Society}, 
  title={Line length: an efficient feature for seizure onset detection}, 
  year={2001},
  volume={2},
  number={},
  pages={1707-1710 vol.2},
  doi={10.1109/IEMBS.2001.1020545}}

@article{ganguly2022seizure,
  title={Seizure detection in continuous inpatient EEG: a comparison of human vs automated review},
  author={Ganguly, Taneeta Mindy and Ellis, Colin A and Tu, Danni and Shinohara, Russell T and Davis, Kathryn A and Litt, Brian and Pathmanathan, Jay},
  journal={Neurology},
  volume={98},
  number={22},
  pages={e2224--e2232},
  year={2022},
  publisher={AAN Enterprises}
}

@article{grant2014eeg,
  title={EEG interpretation reliability and interpreter confidence: a large single-center study},
  author={Grant, Arthur C and Abdel-Baki, Samah G and Weedon, Jeremy and Arnedo, Vanessa and Chari, Geetha and Koziorynska, Ewa and Lushbough, Catherine and Maus, Douglas and McSween, Tresa and Mortati, Katherine A and others},
  journal={Epilepsy \& Behavior},
  volume={32},
  pages={102--107},
  year={2014},
  publisher={Elsevier}
}

@article{huang2023survey,
  title={A survey on hallucination in large language models: Principles, taxonomy, challenges, and open questions},
  author={Huang, Lei and Yu, Weijiang and Ma, Weitao and Zhong, Weihong and Feng, Zhangyin and Wang, Haotian and Chen, Qianglong and Peng, Weihua and Feng, Xiaocheng and Qin, Bing and others},
  journal={arXiv preprint arXiv:2311.05232},
  year={2023}
}

@misc{nye2021work,
      title={Show Your Work: Scratchpads for Intermediate Computation with Language Models}, 
      author={Maxwell Nye and Anders Johan Andreassen and Guy Gur-Ari and Henryk Michalewski and Jacob Austin and David Bieber and David Dohan and Aitor Lewkowycz and Maarten Bosma and David Luan and Charles Sutton and Augustus Odena},
      year={2021},
      eprint={2112.00114},
      archivePrefix={arXiv},
      primaryClass={cs.LG}
}

@article{wang2020fewshot,
    author = {Wang, Yaqing and Yao, Quanming and Kwok, James T. and Ni, Lionel M.},
    title = {Generalizing from a Few Examples: A Survey on Few-Shot Learning},
    year = {2020},
    issue_date = {May 2021},
    publisher = {Association for Computing Machinery},
    address = {New York, NY, USA},
    volume = {53},
    number = {3},
    issn = {0360-0300},
    url = {https://doi.org/10.1145/3386252},
    doi = {10.1145/3386252},
    journal = {ACM Comput. Surv.},
    month = {jun},
    articleno = {63},
    numpages = {34}
}

@article{lopez2017tuab,
  title={Automated Identification of Abnormal EEGs.},
  author={Lopez, S.},
  year={2017},
  organization={Temple University}
}

@misc{li2023cancergpt,
      title={CancerGPT: Few-shot Drug Pair Synergy Prediction using Large Pre-trained Language Models}, 
      author={Tianhao Li and Sandesh Shetty and Advaith Kamath and Ajay Jaiswal and Xianqian Jiang and Ying Ding and Yejin Kim},
      year={2023},
      eprint={2304.10946},
      archivePrefix={arXiv},
      primaryClass={cs.CL}
}

@misc{liu2023large,
      title={Large Language Models are Few-Shot Health Learners}, 
      author={Xin Liu and Daniel McDuff and Geza Kovacs and Isaac Galatzer-Levy and Jacob Sunshine and Jiening Zhan and Ming-Zher Poh and Shun Liao and Paolo Di Achille and Shwetak Patel},
      year={2023},
      eprint={2305.15525},
      archivePrefix={arXiv},
      primaryClass={cs.CL}
}

@misc{lightman2023lets,
      title={Let's Verify Step by Step}, 
      author={Hunter Lightman and Vineet Kosaraju and Yura Burda and Harri Edwards and Bowen Baker and Teddy Lee and Jan Leike and John Schulman and Ilya Sutskever and Karl Cobbe},
      year={2023},
      eprint={2305.20050},
      archivePrefix={arXiv},
      primaryClass={cs.LG}
}

@article{vaswani2017attention,
  title={Attention is all you need},
  author={Vaswani, Ashish and Shazeer, Noam and Parmar, Niki and Uszkoreit, Jakob and Jones, Llion and Gomez, Aidan N and Kaiser, {\L}ukasz and Polosukhin, Illia},
  journal={Advances in neural information processing systems},
  volume={30},
  year={2017}
}

@article{wang2023hypothesis,
  title={Hypothesis search: Inductive reasoning with language models},
  author={Wang, Ruocheng and Zelikman, Eric and Poesia, Gabriel and Pu, Yewen and Haber, Nick and Goodman, Noah D},
  journal={arXiv preprint arXiv:2309.05660},
  year={2023}
}

@misc{yang2023mmreact,
      title={MM-REACT: Prompting ChatGPT for Multimodal Reasoning and Action}, 
      author={Zhengyuan Yang and Linjie Li and Jianfeng Wang and Kevin Lin and Ehsan Azarnasab and Faisal Ahmed and Zicheng Liu and Ce Liu and Michael Zeng and Lijuan Wang},
      year={2023},
      eprint={2303.11381},
      archivePrefix={arXiv},
      primaryClass={cs.CV}
}

@misc{yao2023tree,
      title={Tree of Thoughts: Deliberate Problem Solving with Large Language Models}, 
      author={Shunyu Yao and Dian Yu and Jeffrey Zhao and Izhak Shafran and Thomas L. Griffiths and Yuan Cao and Karthik Narasimhan},
      year={2023},
      eprint={2305.10601},
      archivePrefix={arXiv},
      primaryClass={cs.CL}
}

\end{document}